\documentclass[10pt]{iopart}  
\usepackage{graphicx}
\usepackage{epsf}
\usepackage{amssymb}
\def\Q1{{\bf Q}_1}

\def\e{{\rm e}}

\def\gs{\gamma_S}

\def\ls{\lambda_S}
\def\vexp{$VT^3 \, \exp[-0.7 \, {\rm GeV}/T]$}

\def\ssb{\langle {\rm s}\bar{\rm s}\rangle}
\def\uub{\langle {\rm u}\bar{\rm u}\rangle}
\def\ddb{\langle {\rm d}\bar{\rm d}\rangle}

\def\NP{\mathrm{N}_\mathrm{P}}

\def\np{$N_\mathrm{P}$ }

\newcommand{\NN}{NN }

\begin{document}

\title[Chemical equilibrium study at SPS 158$A$ GeV]{Chemical equilibrium study at SPS 158$A$ GeV}

\author{Jaakko Manninen \dag\
\footnote[3]{jaakkoma@paju.oulu.fi}
}
\address{\dag\ University of Oulu, Oulu, Finland}

\begin{abstract}
A detailed study of chemical freeze-out in nucleus-nucleus collisions 
at beam energy 158$A$ GeV is presented. By analyzing hadronic 
multiplicities within the statistical hadronization approach,
the chemical equilibration of\\ p-p, C-C, Si-Si and Pb-Pb systems is studied as 
a function of the number of participating nucleons in the system.
Additionally, the Two Component statistical hadronization model is applied to the data
and is found to be able to explain the observed strangeness hadronic phase space 
under-saturation.
\end{abstract}

%\keyword{heavy ion, quark gluon plasma, hadronization, statistical model} 
%\PACS{specifications see, e.g.\ {\tt http://www.aip.org/pacs/}}
%24.10.Pa=Thermal and statistical models
%25.75.Dw=Particle and resonance production 


%\maketitle

%************************************************************************
\section{The statistical hadronization model}
%************************************************************************

The main idea of the SHM is that hadrons are emitted from regions at statistical 
equilibrium. No hypothesis is made about how 
statistical equilibrium is achieved; this can be a direct consequence of the 
hadronization process~\cite{beca_bari}. In a single collision event, there might 
be several clusters with different collective momenta, different
overall charges and volumes. However, Lorentz-invariant quantities like particle 
multiplicities are independent of clusters' momenta.
 
Depending on the system size, statistical analysis can be done in different
statistical ensembles, which have been studied 
carefully by different groups and methods suitable for relativistic nucleus-nucleus 
collisions are well established~\cite{kerabeca}.
In this work, small systems with net baryon number less than 10 are 
calculated in the $BSQ$-canonical ensemble taking into account exact conservation of
$B$, $S$ and $Q$ charges. For the systems with the number of participants ($N_\mathrm{P}$) 
between 10 and 100, exact conservation of
strangeness only is taken into account while larger systems are treated 
grand-canonically (GC).

In spite of the appropriate ensemble, theoretical multiplicities are calculated within the 
main version 
of the statistical model by fitting temperature $T$, scaling volume $V$, 
and strangeness suppression factor $\gamma_S$~\cite{gammas}. 
Additionally, one needs to introduce a chemical potential for all charges that are
treated in a GC manner. Usually the baryon chemical potential is taken as a free fit parameter,  
while the 
strangeness chemical potential is fixed by assuming net strangeness neutrality 
in the system and the chemical potential for electric charge by assuming that $Q/B$ in the 
system equals $Z/A$ of the colliding nuclei.

The overall
multiplicity to be compared with the data, is calculated as the sum of primary 
multiplicity and the contribution from the decay of heavier hadrons:
$\langle n_j \rangle  = \langle n_j \rangle^{\mathrm{primary}} + 
\sum_k \mathrm{Br}(k\rightarrow j) \langle n_k \rangle$,
where the branching ratios are taken from the latest issue of the Review of Particle 
Physics \cite{pdg}.

%\begin{table}[!ht]
%\begin{footnotesize}
%\vspace{0.3cm}
%\begin{tabular}{|c|c|c|c|c|}
%\hline
% Parameters     & Main analysis       &  SHM(TC)        &   Main analysis       &  SHM(TC)  \\
%\hline               
% & \multicolumn{2}{|c|}{C-C 158$A$ GeV} &  \multicolumn{2}{|c|}{Si-Si 158$A$ GeV} \\
%\hline               
%$T$ (MeV)             & 165.7$\pm$4.1 (4.1)         &   170$\pm$10 (13) &    163.0$\pm$4.7 (6.5)     &   162.0$\pm$7.6 (7.6) \\
%$\mu_B$ (MeV)         & 248.1$\pm$12.5 (12.5)       &               &    245.5$\pm$11.0 (15.1)   &   234.4$\pm$22.5 (22.5)\\
%$\gs$                 & 0.575$\pm$0.042 (0.042)     &   1.0 (fixed) &    0.664$\pm$0.050 (0.069) &   1.0 (fixed)  \\
% V'                   & 0.84$\pm$0.05  (0.05)       &  0.23$\pm$0.03 (0.04) &    2.10$\pm$0.13 (0.18)    &  0.91$\pm$0.11 (0.11) \\
%$\langle N_c \rangle$ &                             &   6$\pm$0.4 (0.4)   &                            &   11.4$\pm$1.8 (1.8) \\
%\hline
%$\chi^2$/dof          & 3.4/4                       & 5.8/5         &    7.6/4                   & 1.0/4\\
%\hline
%\end{tabular}
%\end{footnotesize}
%\caption{Summary of fitted parameters (V'=\vexp) at top SPS beam energy in 
%the framework of the SHM($\gamma_S$) model and SHM(TC) .
%The re-scaled errors (see~\cite{pdg,PbPb_last}) are quoted within brackets.
%\vspace{-0.2cm}\label{parameters}}
%\end{table}

\begin{table}[!ht]
\begin{footnotesize}
\vspace{0.3cm}
\begin{tabular}{|c|c|c|c|c|c|}
\hline
 Parameters     &  SHM($\gamma_S$)  &  SHM($\gamma_S$)       &  SHM(TC)        &  SHM($\gamma_S$)       &  SHM(TC)  \\
\hline               
 & p-p 158$A$ GeV    & \multicolumn{2}{|c|}{C-C 158$A$ GeV} &  \multicolumn{2}{|c|}{Si-Si 158$A$ GeV} \\
\hline               
$T$ (MeV)             & 177.3$\pm$5.2     & 165.7$\pm$4.1         &   170$\pm$10  &    163.0$\pm$4.7      &   162.0$\pm$7.6 \\
$\mu_B$ (MeV)         &                         & 248.1$\pm$12.5       &               &    245.5$\pm$11.0   &   234.4$\pm$22.5 \\
$\gs$                 & 0.445$\pm$0.020 & 0.575$\pm$0.042     &   1.0 (fixed) &    0.664$\pm$0.050  &   1.0 (fixed)  \\
 V'                   &  0.128$\pm$0.005 & 0.84$\pm$0.05        &  0.23$\pm$0.03  &    2.10$\pm$0.13    &  0.91$\pm$0.11  \\
$\langle N_c \rangle$ & &                             &   6$\pm$0.4   &                            &   11.4$\pm$1.8 \\
\hline
$\chi^2$/dof          & 13.0/7 & 3.4/4                       & 5.8/5         &    7.6/4                   & 1.0/4\\
\hline
\end{tabular}
\end{footnotesize}
\caption{Summary of fitted parameters (V'=\vexp) at top SPS beam energy in 
the framework of the SHM($\gamma_S$) model and SHM(TC).
\vspace{-0.2cm}\label{parameters}}
\end{table}

%************************************************************************
\section{Experimental data set and analysis results}
%************************************************************************

\begin{figure}[!ht]
$\begin{array}{c@{\hspace{0.3in}}c}
\epsfxsize=3.5in
\epsffile{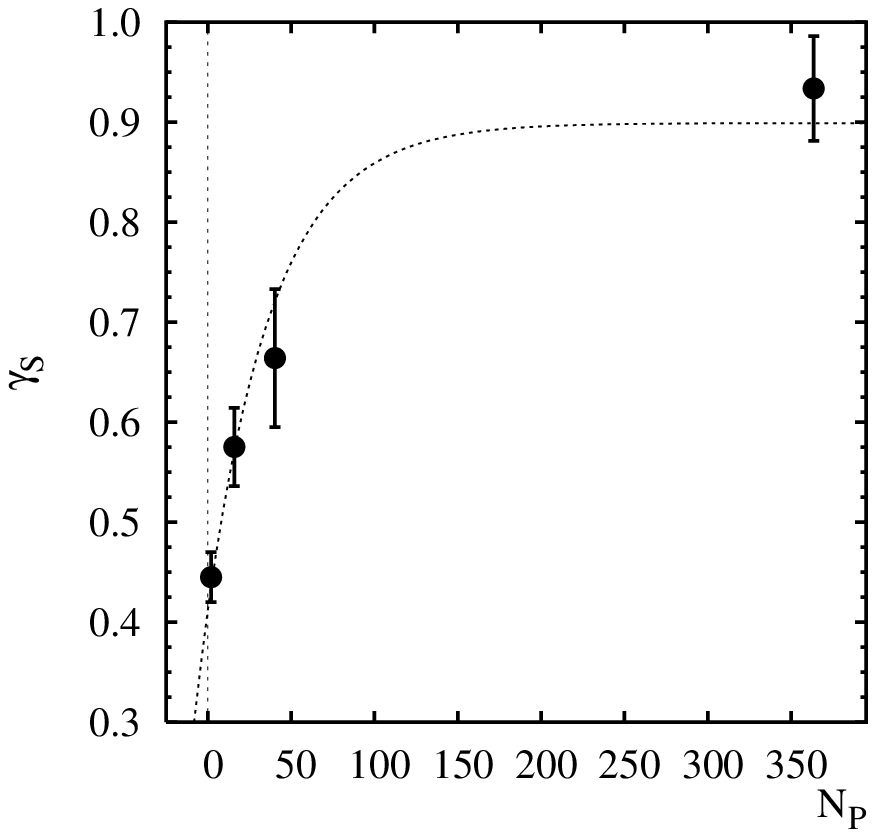} & \hspace*{-3cm}
	\epsfxsize=3.5in
	\epsffile{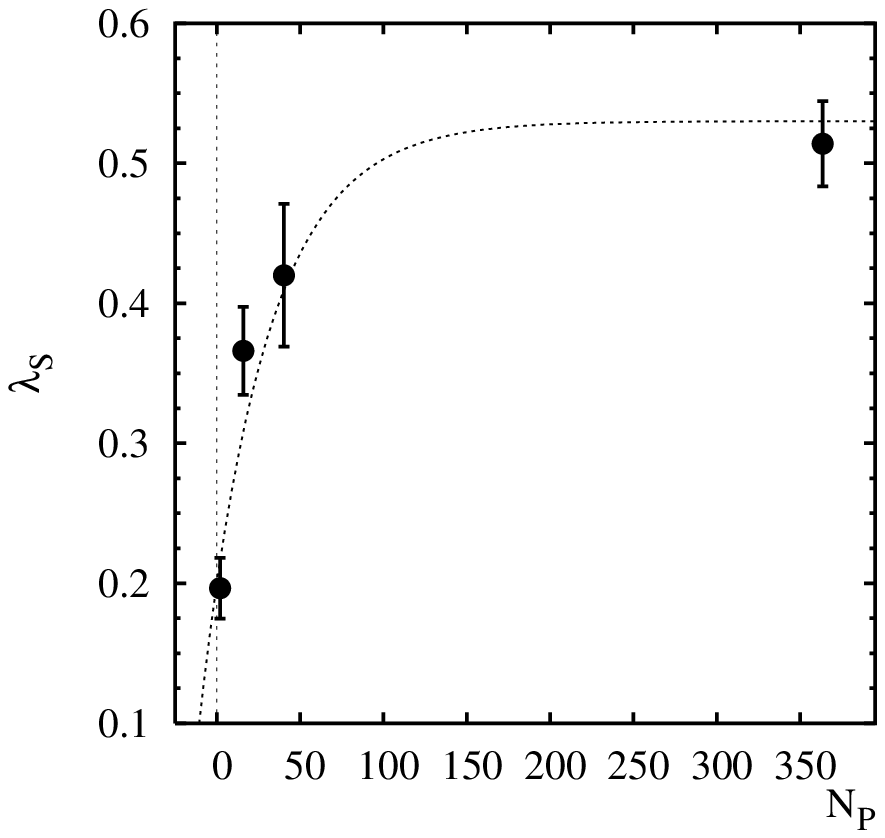} \\
\end{array}$
\caption{The strangeness non-equilibrium parameter $\gamma_S$ and the corresponding 
 $\ls$ factor as a function of the number of participating nucleons in the collision system
(Pb-Pb from~\cite{PbPb_last}). 
Both lines are of the functional form
$f(\NP)=A-\e^{-B\NP}$ and are plotted to guide the eye. 
Whenever the resulting $\chi^2/dof > 1$ in the fit, errors have been re-scaled by factor 
$\sqrt{\chi^2/dof}$, for details see~\cite{pdg,PbPb_last}. 
From left to right: 
p-p, C-C, Si-Si and Pb-Pb.\label{gsls}}
\end{figure}

The experimental data consists of measurements made by the NA49 
collaboration in central p-p, C-C, Si-Si and Pb-Pb collisions at beam momenta 
158$A$ GeV~\cite{NA49pp_all,prel_CCSiSi,Af2002mx,Mischke2003,Anticic2004,Friese2002,Alt2004}.  
The analysis has been carried out by looking for the minimum of the 
$\chi^2 = \sum_i \frac{(n_i^{\rm exp} - n_i^{\rm theo})^2}{\sigma_i^2}$.
The fitted parameters within the main scheme SHM($\gs$) are shown in Table~\ref{parameters}.
A major result of these fits is that $\gs$ is monotonically increasing
as a function of the number of participating nucleons, 
and significantly smaller than 1 in all cases (see Fig.\ref{gsls}), 
thus strangeness seems to be under-saturated with respect 
to a completely chemically equilibrated hadron gas. This confirms previous 
findings \cite{PbPb_last,beca01,cley,cley_systemsize}.

%==================================================================================
\subsection{{\bf Superposition of \NN collisions with an equilibrated fireball}}
%==================================================================================

In the Two Component model (SHM(TC)), first introduced in~\cite{PbPb_last}, the observed 
hadron production is taken as the superposition of two components: one originated from a large 
fireball at complete chemical equilibrium at freeze-out, with $\gs=1$, and another 
component from single nucleon-nucleon collisions. 
Since it is known that in \NN collisions strangeness is strongly suppressed, %~\cite{beca2},
the idea is to ascribe the observed under-saturation of strangeness in heavy
ion collisions to the \NN component.

With the simplifying assumption of disregarding 
subsequent inelastic collisions of particles produced in those primary \NN collisions, 
the overall hadron multiplicity can be written then as 
$ \langle n_j \rangle = \langle N_c \rangle \langle n_j \rangle_{NN} +
 \langle n_j \rangle_V,$
where $\langle n_j \rangle_{NN}$ is the average multiplicity of the $j^{\rm th}$
hadron in a single \NN collision, $\langle N_c \rangle$ is the mean number of
single \NN collisions and $\langle n_j 
\rangle_V$ is the average multiplicity of hadrons emitted from the equilibrated
fireball.
To estimate $\langle n_j \rangle_{NN}$ in np and nn collisions, the 
parameters (see Table \ref{parameters}) of the statistical model determined in pp 
are retained and the 
initial quantum numbers are changed accordingly. Theoretical multiplicities have 
been calculated in the canonical ensemble, which is described in detail in 
ref.~\cite{becapt}. 

This model was seen to be able to reproduce the experimental particle multiplicities
measured in the Pb-Pb collisions at 158$A$ GeV~\cite{PbPb_last,jamaica_proc}.
For the Si-Si system, $T$, $V$, $\mu_B$ of the central fireball and $\langle N_c \rangle$ 
were fitted using  the S-canonical ensemble in the central fireball.
The fit quality is significantly improved compared to the main version of the statistical 
model, if  the number of
"single" \NN collisions is about 11 with a 16\% uncertainty.

The central fireball produced in the C-C system needs to be analyzed in the 
$BSQ$-canonical ensemble
taking into account the actual proton-neutron configurations. The Two Component model can 
be used to describe the C-C system as well, if the total baryon number in the central 
fireball is $B=N_p+N_n\approx4$. Even though fit quality is slightly worse than with the 
main version of the statistical model, experimental multiplicities are well reproduced
with the SHM(TC) also.

%************************************************************************
\section{System size dependence and conclusions} 
%************************************************************************

Our fit results show non-trivial system size dependence of chemical equilibration
and characteristic thermal parameters of the source in ultra-relativistic heavy ion
collisions. 
The chemical freeze-out of all the different 
systems, C-C, Si-Si and Pb-Pb with beam momenta 158$A$ GeV seem to happen at similar 
chemical state, all of them at $\mu_B\approx$250 MeV. Such weak system size dependence of
the baryon chemical potential has been already reported earlier~\cite{cley_systemsize}.

\begin{figure}[!ht]
\begin{center}
\includegraphics[scale=0.9,angle=0]{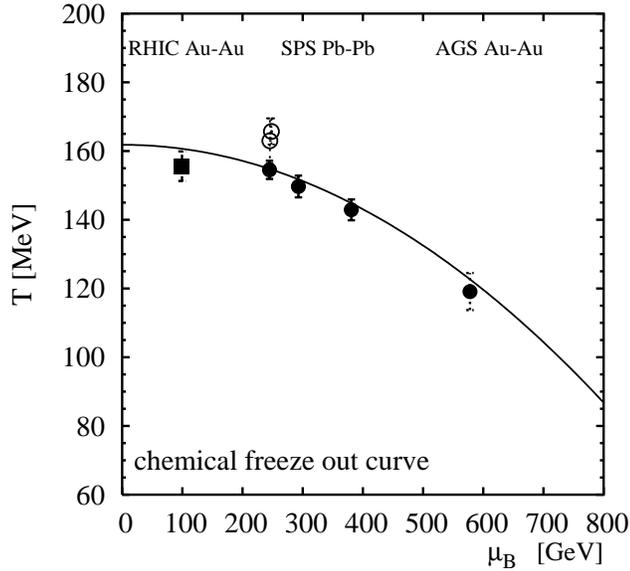}
\end{center}
\caption{Chemical freeze-out points in the [$\mu_B-T$] plane in various heavy ion 
collisions. Full round dots~\cite{PbPb_last} refer to the Au-Au at 11.6 and 
Pb-Pb collisions at 40, 80, 158$A$ GeV, whilst the square dot 
has been obtained by applying the statistical model to the preliminary %4$\pi$ 
$\pi^+$, $\pi^-$, $K^+$, $K^-$ and \np multiplicities~\cite{dieterSQM2004}
in the Au-Au collisions at $\sqrt s_{NN} = 200$ GeV.
The hollow round dots refer to the C-C and Si-Si collisions at 158$A$ GeV.
Whenever the resulting $\chi^2/dof > 1$ in the fit, errors have been re-scaled by factor 
$\sqrt{\chi^2/dof}$, for details see~\cite{pdg,PbPb_last}.
}\label{tmu_trho}
\end{figure}

Systems with few participating 
nucleons seem to decouple at slightly higher temperature than heavy systems, but in general 
the chemical freeze-out temperature as well as the baryon chemical potential are
determined mostly by the beam energy, not by the number of participants, and thus
C-C, Si-Si and Pb-Pb with the same beam momenta, seem not
to follow the chemical freeze-out curve (see Fig.~\ref{tmu_trho}), but show more 
complex system size dependence,
an interplay of the initial beam energy and the number of participants in the system. 

The chemical equilibration of strangeness, on the other hand, seems to be strongly dependent
on the number of participants.
Going from small to large systems, the $\gamma_S$ parameter increases monotonically 
from 0.45 in the p-p to
0.9 in the Pb-Pb with the same beam momenta, and thus strangeness seems to 
be out of equilibrium in all collision systems studied at SPS. The Wroblewski variable 
$\ls = 2 \ssb/(\uub+\ddb)$, the estimated 
ratio of newly produced strange quarks to u and d quarks at primary hadron level,  
features very similar suppression in strangeness production, see Fig.~\ref{gsls}.

% =======================================================================================

\end{document}